\documentclass[12pt]{article}
\usepackage{amssymb}
\usepackage{amsmath}
\usepackage{amsfonts}
\usepackage{epsfig}
\usepackage{graphicx}

\newtheorem{theorem}{Theorem}

\begin{document}

\title{Measure and mass gap for generalized connections on hypercubic
lattices}
\author{R. Vilela Mendes \thanks{%
CMAF, Universidade de Lisboa, Av. Gama Pinto, 2 - 1649-003 Lisboa
(Portugal), rvmendes@fc.ul.pt, rvilela.mendes@gmail.com}}
\date{}
\maketitle

\begin{abstract}
Using projective limits as subsets of Cartesian products of homomorphisms
from a lattice to the structure group, a consistent interaction measure and
an infinite-dimensional calculus has been constructed for a theory of
non-abelian generalized connections on a hypercubic lattice. Here, after
reviewing and clarifying past work, new results are obtained for the mass
gap when the structure group is compact.
\end{abstract}

\section{Introduction}

In \cite{VilelaJMP} a space for generalized connections was defined using
projective limits as subsets of Cartesian products of homomorphisms from a
lattice to a structure group. In this space, non-interacting and interacting
measures were defined as well as functions and operators. From projective
limits of test functions and distributions on products of compact groups, a
projective gauge triplet was obtained, which provides a framework for an
infinite-dimensional calculus in gauge theories.

In \cite{VilelaJMP} a central role is played by the construction of an
interacting measure which, satisfying a consistency condition, can be
extended to a projective limit of decreasing lattice spacing and
increasingly larger lattices. Since \cite{VilelaJMP} was published some
questions have been raised concerning in particular the construction of the
measure and the consistency condition. The purpose of this paper is twofold.
First to clarify and extend some details of the measure construction which,
of course, were implicit in \cite{VilelaJMP}. Second to further explore some
of the physical consequences of the constructed measure, in particular the
nature of the mass gap that it implies.

The basic setting, as used in \cite{VilelaJMP}, is the following:

In $\mathbb{R}^{4}$a sequence of hypercubic lattices is constructed in such
a way that any plaquette of edge size $\frac{a}{2^{k}}$ $\left(
k=0,1,2,\cdots \right) $\ is a refinement of a plaquette of edge $\frac{a}{%
2^{k-1}}$ (meaning that all vertices of the $\frac{a}{2^{k-1}}$ plaquette
are also vertices in the $\frac{a}{2^{k}}$ plaquettes). The refinement is
made one-plaquette-at-a-time, in the sense that, when one plaquette of edge $%
\frac{a}{2^{k-1}}$ is converted into four plaquettes of edge $\frac{a}{2^{k}}
$, eight new plaquettes of edge $\frac{a}{2^{k-1}}$, orthogonal to the
refined plaquette, are also added to the lattice. The additional plaquettes
connect the new vertices of the refined $\frac{a}{2^{k}}$ plaquette to the
middle points of $\frac{a}{2^{k-1}}$ plaquettes, in such a way that when all 
$\frac{a}{2^{k-1}}$ plaquettes are refined to $\frac{a}{2^{k}}$ size, a full
hypercubic $\frac{a}{2^{k}}$ lattice is obtained. See Fig.1 for a $3-$%
dimensional projection of the process, where two of the additional eight (in 
$\mathbb{R}^{4}$) plaquettes are shown, attached to the points $A,B,C$ and $%
D $. This one-plaquette-at-a-time construction is useful to check the
consistency condition (see Section 2).

Finite volume hypercubes $\Gamma $ in these lattices form a directed set $%
\left\{ \Gamma ,\succ \right\} $ under the inclusion relation $\succ $. $%
\Gamma \succ \Gamma ^{\prime }$ meaning that all edges and vertices in $%
\Gamma ^{\prime }$ are contained in $\Gamma $, the inclusion relation
satisfying%
\begin{eqnarray}
\Gamma &\succ &\Gamma  \notag \\
\Gamma &\succ &\Gamma ^{\prime }\text{ and }\Gamma ^{\prime }\succ \Gamma
\Longrightarrow \Gamma =\Gamma ^{\prime }  \notag \\
\Gamma &\succ &\Gamma ^{\prime }\text{ and }\Gamma ^{\prime }\succ \Gamma
^{\prime \prime }\Longrightarrow \Gamma \succ \Gamma ^{\prime \prime }
\label{I.1}
\end{eqnarray}%
After each complete refinement of a finite volume hypercube (from $\frac{a}{%
2^{k-1}}$ to $\frac{a}{2^{k}}$ size), the sequence is expanded to include
larger and larger volume hypercubes which are likewise refined, etc..

\begin{figure}[htb]
\centering
\includegraphics[width=0.8\textwidth]{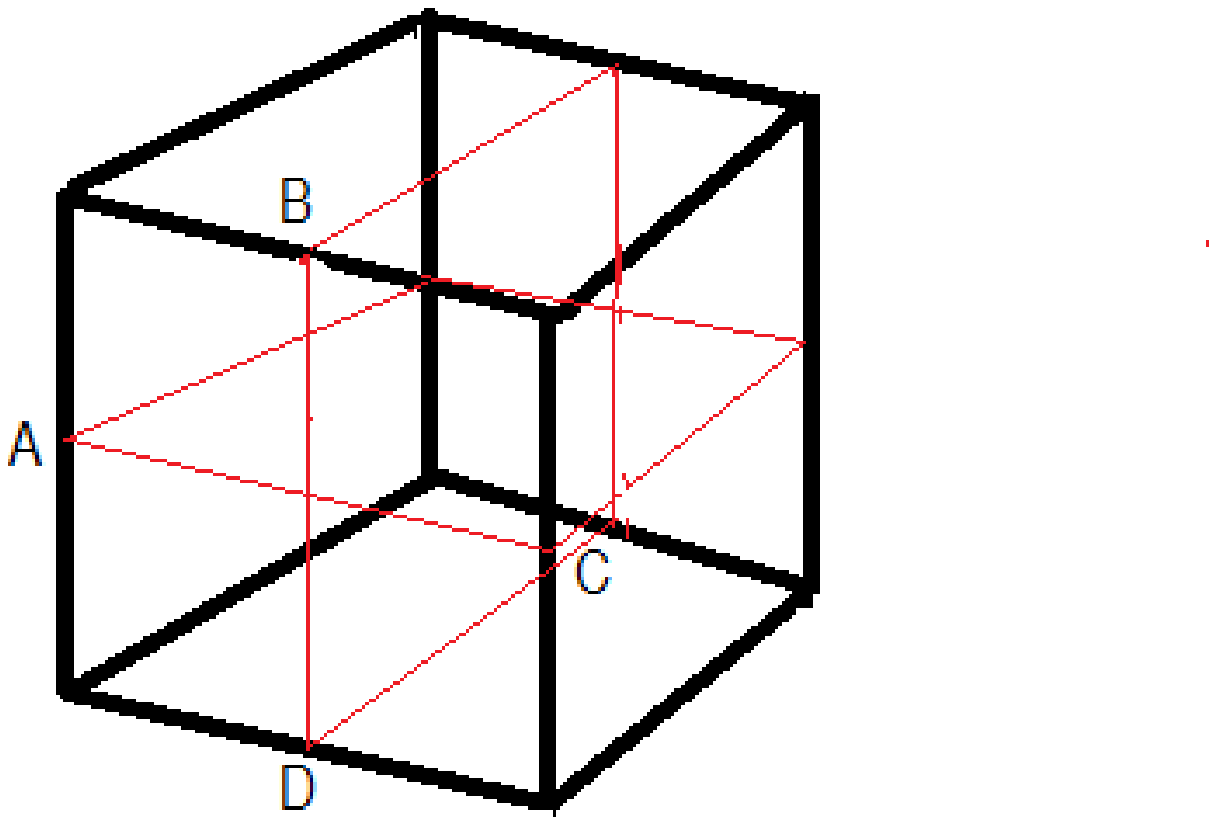}
\caption{Partial 3-dimensional projection of the one-plaquette-at-a-time
refinement process}
\end{figure}

Let $\mathbb{G}$ be a compact group and $x_{0}$ a point that does not belong
to any lattice point of the directed family. Assuming an analytic
parametrization of each edge, associate to each edge $l$ a $x_{0}$-based
loop and for each generalized connection $A$ consider the holonomy $%
h_{l}\left( A\right) $ associated to this loop. For definiteness each edge
is considered to be oriented along the coordinates positive direction and
the set of edges of the lattice $\Gamma $ is denoted $E\left( \Gamma \right) 
$. The set $\mathcal{A}_{\Gamma }$ of generalized connections for the
lattice hypercube $\Gamma $ is the set of homomorphisms $\mathcal{A}_{\Gamma
}=Hom\left( E\left( \Gamma \right) ,G\right) \sim G^{\#E\left( \Gamma
\right) }$, obtained by associating to each edge the holonomy $h_{l}\left(
\cdot \right) $ on the associated $x_{0}$-based loop. The set of
gauge-independent generalized connections $\mathcal{A}_{\Gamma }/Ad$ is
obtained factoring by the adjoint representation at $p_{0}$, $\mathcal{A}%
_{\Gamma }/Ad$ $\sim G^{\#E\left( \Gamma \right) }/Ad$. However because, for
gauge independent functions, integration in $\mathcal{A}_{\Gamma }$
coincides with integration in $\mathcal{A}_{\Gamma }/Ad$ , for simplicity,
from now on one uses only $\mathcal{A}_{\Gamma }$. Finally one considers the
projective limit $\mathcal{A}=\underset{\longleftarrow }{\lim }\mathcal{A}%
_{\Gamma }$ of the family 
\begin{equation}
\left\{ \mathcal{A}_{\Gamma },\pi _{\Gamma \Gamma ^{\prime }}:\Gamma
^{\prime }\succ \Gamma \right\}  \label{I.2}
\end{equation}%
$\pi _{\Gamma \Gamma ^{\prime }}$ and $\pi _{\Gamma }$ denoting the
surjective projections $\mathcal{A}_{\Gamma ^{\prime }}\longrightarrow 
\mathcal{A}_{\Gamma }$ and $\mathcal{A}\longrightarrow \mathcal{A}_{\Gamma }$%
.

The projective limit of the family $\left\{ \mathcal{A}_{\Gamma },\pi
_{\Gamma \Gamma ^{\prime }}\right\} $ is the subset $\mathcal{A}$ of the
Cartesian product $\underset{\Gamma }{\prod }\mathcal{A}_{\Gamma }$ defined
by 
\begin{equation}
\mathcal{A}=\left\{ a\in \underset{\Gamma }{\prod }\mathcal{A}_{\Gamma
}:\Gamma ^{\prime }\succ \Gamma \Longrightarrow \pi _{\Gamma \Gamma ^{\prime
}}\mathcal{A}_{\Gamma ^{^{\prime }}}=\mathcal{A}_{\Gamma }\right\}
\label{I.3}
\end{equation}%
the projective topology in $\mathcal{A}$ being the coarsest topology for
which each $\pi _{\Gamma }$ mapping is continuous.

For a compact group $\mathbb{G}$, each $\mathcal{A}_{\Gamma }$ is a compact
Hausdorff space. Then $\mathcal{A}$ is also a compact Hausdorff space. In
each $\mathcal{A}_{\Gamma }$ one has a natural (Haar) normalized product
measure $\nu _{\Gamma }=\mu _{H}^{\#E\left( \Gamma \right) }$, $\mu _{H}$
being the normalized Haar measure in $\mathbb{G}$. Then, according to a
theorem of Prokhorov, as generalized by Kisynski \cite{Kisynski} \cite%
{Maurin}, if 
\begin{equation}
\nu _{\Gamma ^{\prime }}\left( \pi _{\Gamma \Gamma ^{\prime }}^{-1}\left(
B\right) \right) =\nu _{\Gamma }\left( B\right)  \label{I.4}
\end{equation}%
for every $\Gamma ^{\prime }\succ \Gamma $ and every Borel set $B$ in $%
\mathcal{A}_{\Gamma }$, there is a unique measure $\nu $ in $\mathcal{A}$
such that $\nu \left( \pi _{\Gamma }^{-1}\left( B\right) \right) =\nu
_{\Gamma }\left( B\right) $ for every $\Gamma $.

\section{The measure}

As stated before, the essential step in the construction of the measure in
the projective limit is the fulfilling of the consistency condition (\ref%
{I.4}). One considers, on the finite-dimensional spaces $\mathcal{A}_{\Gamma
}\sim G^{\#E\left( \Gamma \right) }$, measures that are absolutely
continuous with respect to the Haar measure 
\begin{equation}
d\mu _{\mathcal{A}_{\Gamma }}=p\left( \mathcal{A}_{\Gamma }\right) \left(
d\mu _{H}\right) ^{\#E\left( \Gamma \right) }  \label{II.1}
\end{equation}%
$p\left( \mathcal{A}_{\Gamma }\right) $ being a continuous function in $%
\mathcal{A}_{\Gamma }$ with the simplifying assumptions:

- $p\left( \mathcal{A}_{\Gamma }\right) $ is a product of plaquette
functions 
\begin{equation}
p\left( \mathcal{A}_{\Gamma }\right) =p\left( U_{\square _{1}}\right)
p\left( U_{\square _{2}}\right) \cdots p\left( U_{\square _{n}}\right)
\label{II.2}
\end{equation}%
with $U_{\square }\left( A_{\Gamma }\right) =h_{1}h_{2}h_{3}^{-1}h_{4}^{-1}$%
, $h_{1}$ to $h_{4}$ being the holonomies of the $x_{0}-$ based loops
associated to the edges of the plaquette.

- $p\left( \cdot \right) $ is a central function, $p\left( xy\right)
=p\left( yx\right) $ or, equivalently $p\left( y^{-1}xy\right) =p\left(
x\right) $ with $x,y\in \mathbb{G}$.

Let $p^{\prime },p^{\prime \prime }$ and $p$ be the density functions
associated respectively to the square plaquette with edges of size $\frac{a}{%
2^{k}}$, to the rectangular plaquette with edges of size $\frac{a}{2^{k}}$
and $\frac{a}{2^{k-1}}$ and, finally, to the square plaquette with edges of
size $\frac{a}{2^{k-1}}$. Then

\begin{theorem}
\cite{VilelaJMP} \textit{A measure on the projective limit }$\mathcal{A}=%
\underset{\longleftarrow }{\lim }\mathcal{A}_{\Gamma }$\textit{\ exists if a
sequence of functions is found satisfying }%
\begin{eqnarray}
\int p^{\prime }\left( G_{i}X\right) p^{\prime }\left( X^{-1}G_{j}\right)
d\mu _{H}\left( X\right) &\sim &p^{\prime \prime }\left( G_{i}G_{j}\right) 
\notag \\
\int p^{\prime \prime }\left( G_{i}X\right) p^{\prime \prime }\left(
X^{-1}G_{j}\right) d\mu _{H}\left( X\right) &\sim &p\left( G_{i}G_{j}\right)
\label{II.3}
\end{eqnarray}%
\textit{for plaquette subdivisions of all sizes.}
\end{theorem}

\textbf{Proof:} In the directed set $\left\{ \Gamma ,\succ \right\} $
consider two elements $\Gamma $ and $\Gamma ^{^{\prime }}$ which differ only
in subdivision of a single plaquette from $\frac{a}{2^{k-1}}$ to $\frac{a}{%
2^{k}}$ size (see Fig.2) plus the additional $\frac{a}{2^{k}}$ plaquettes as
explained in the introduction.

\begin{figure}[htb]
\centering
\includegraphics[width=0.8\textwidth]{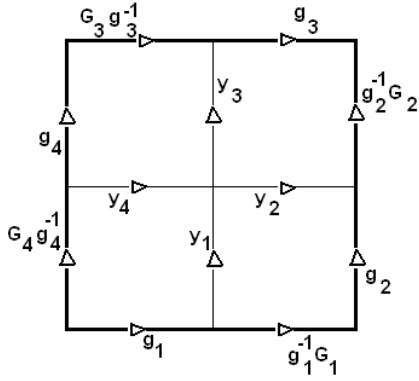}
\caption{Subdivision of one plaquette}
\end{figure}

The consistency condition is 
\begin{eqnarray}
&&\frac{1}{Z^{^{\prime }}}\int p^{\prime }\left(
g_{1}^{-1}G_{1}g_{2}y_{2}^{-1}y_{1}^{-1}\right) p^{\prime }\left(
y_{2}g_{2}^{-1}G_{2}g_{3}^{-1}y_{3}^{-1}\right) p^{\prime }\left(
y_{4}y_{3}g_{3}G_{3}^{-1}g_{4}^{-1}\right) p^{\prime }\left(
g_{1}y_{1}y_{4}^{-1}g_{4}G_{4}^{-1}\right)  \notag \\
&&\prod_{i=1}^{4}d\mu _{H}\left( g_{i}\right) d\mu _{H}\left( y_{i}\right)
d\mu _{H}\left( G_{i}\right) \prod_{k=1}^{8}d\mu _{H}\left( G_{k}\right) 
\notag \\
&=&\frac{1}{Z}\int p\left( G_{1}G_{2}G_{3}^{-1}G_{4}^{-1}\right)
\prod_{i=1}^{4}d\mu _{H}\left( G_{i}\right)  \label{II.4}
\end{eqnarray}%
the last factor in the left hand side denoting integration over the
additional $\frac{a}{2^{k}}$ plaquettes. Using centrality of $p^{\prime }$,
redefining 
\begin{equation}
g_{1}y_{1}=X_{1},\qquad g_{2}y_{2}^{-1}=X_{2},\qquad
y_{3}g_{3}=X_{3}^{-1},\qquad y_{4}^{-1}g_{4}=X_{4}^{-1}  \label{II.5}
\end{equation}%
and using invariance of the normalized Haar measure, one may integrate over $%
y_{1},y_{2},y_{3},y_{4}$ and $G_{k}$, obtaining for the left hand side of (%
\ref{II.4}) 
\begin{equation*}
\frac{1}{Z^{^{\prime }}}\int p^{\prime }\left( X_{1}^{-1}G_{1}X_{2}\right)
p^{\prime }\left( X_{2}^{-1}G_{2}X_{3}\right) p^{\prime }\left(
X_{3}^{-1}G_{3}^{-1}X_{4}\right) p^{\prime }\left(
X_{4}^{-1}G_{4}^{-1}X_{1}\right) \prod_{i=1}^{4}d\mu _{H}\left( X_{i}\right)
d\mu _{H}\left( G_{i}\right)
\end{equation*}%
Therefore if there is a sequence of central functions $p^{\prime },p^{\prime
\prime },p$ satisfying the proportionality relations 
\begin{eqnarray}
\int p^{\prime }\left( G_{i}X\right) p^{\prime }\left( X^{-1}G_{j}\right)
d\mu _{H}\left( X\right) &\sim &p^{\prime \prime }\left( G_{i}G_{j}\right) 
\notag \\
\int p^{\prime \prime }\left( G_{i}X\right) p^{\prime \prime }\left(
X^{-1}G_{j}\right) d\mu _{H}\left( X\right) &\sim &p\left( G_{i}G_{j}\right)
\label{II.6}
\end{eqnarray}%
the consistency condition (\ref{II.4}) would be satisfied, with the
proportionality constant absorbed in the overall measure normalization. Then
a measure would exist in the projective limit, because all elements in the
directed set $\left\{ \Gamma ,\succ \right\} $ may be reached by
one-plaquette subdivisions.

If $p\left( U_{\square }\right) $ is a constant, $d\mu _{\mathcal{A}_{\Gamma
}}$ is factorizable and the consistency condition is trivially satified. $%
d\mu _{\mathcal{A}_{\Gamma }}$ would be the Ashtekar-Lewandowski measure for
generalized connections \cite{Ashtekar2} \cite{Ashtekar3}. A nontrivial
solution that satisfies the consistency condition (\ref{II.4}) is the choice
of $p\left( U_{\square }\right) $ as the heat kernel 
\begin{equation}
K\left( g,\beta \right) =\sum_{\lambda \in \Lambda ^{+}}d_{\lambda
}e^{-c\left( \lambda \right) \beta }\chi _{\lambda }\left( g\right)
\label{II.7}
\end{equation}%
with%
\begin{eqnarray}
\beta &\rightarrow &\beta ^{\prime }=\frac{\beta }{4}  \notag \\
\beta &\rightarrow &\beta ^{\prime \prime }=\frac{\beta }{2}  \label{II.8}
\end{eqnarray}%
$\beta ^{\prime \prime },\beta ^{\prime }$ and $\beta $ being the constants
associated to $p^{\prime \prime },p^{\prime }$ and $p$. In (\ref{II.7}), $%
g\in \mathbb{G}$, $\beta \in \mathbb{R}^{+}$, $\Lambda ^{+}$ is the set of
highest weights, $d_{\lambda }$ and $\chi _{\lambda }\left( \cdot \right) $
the dimension and character of the $\lambda -$representation and $c\left(
\lambda \right) $ the spectrum of the Laplacian $\Delta
_{G}:=\sum_{i=1}^{n}\chi _{i}^{2}$, $\left\{ \chi _{i}\right\} $ being a
basis for the Lie algebra of $\mathbb{G}$.

Finally, one writes for the measure on the lattice $\Gamma $%
\begin{equation}
d\mu _{\mathcal{A}_{\Gamma }}=\frac{1}{Z_{\Gamma }}\prod_{edges}d\mu
_{H}(g_{l})\prod_{plaquettes}\sum_{\lambda \in \Lambda ^{+}}d_{\lambda
}e^{-c\left( \lambda \right) \beta }\chi _{\lambda }\left( g_{p}\right)
\label{II.9}
\end{equation}%
and the consistency condition (\ref{I.4}) being satisfied, a measure is also
defined on the projective limit lattice, that is, on the projective limit
generalized connections $\mathcal{A}$.

This measure has the required naive continuum limit, both for abelian and
non-abelian theories (see \cite{VilelaJMP}). Furthermore by defining
infinite-dimensional test functionals and distributions, a projective
triplet was constructed which provides a framework to develop an
infinite-dimensional calculus over the hypercubical lattice. In particular,
this step is necessary to give a meaning to the density $p\left( \mathcal{A}%
_{\Gamma }\right) $ in the $\beta \rightarrow 0$ limit, where $p\left( 
\mathcal{A}_{\Gamma }\right) $ would no longer be a continuous function.
Thus $p\left( \mathcal{A}_{\Gamma }\right) $, a density that multiplies the
Ashtekar-Lewandowski measure \cite{Ashtekar2} \cite{Ashtekar3} \cite%
{Fleischhack3}, gains a distributional meaning in the framework of the
projective triplet.

A theory being completely determined whenever its measure is specified, the
construction in \cite{VilelaJMP} provides a rigorous specification of a
projective limit Yang-Mils theory for gauge fields over a compact group.
Some of the consequences of this specification were already discussed in 
\cite{VilelaJMP}. Here one analyses the nature of the mass gap which follows
from the measure specification.

\section{The mass gap}

The experimental phenomenology of subnuclear physics provides evidence for
the short range of strong interactions. Therefore, if unbroken non-abelian
Yang-Mills is the theory of strong interactions, the Hamiltonian, associated
to its measure, should have a positive mass gap. This important physical
question has been addressed in different ways by several authors. An
interesting research approach \cite{Laufer1} \cite{Laufer2} considers the
Riemannian geometry of the (lattice) gauge-orbit space to compute the Ricci
curvature. The basic inspiration for this approach is the Bochner-Lichn\'{e}%
rowicz \cite{Bochner} \cite{Lichne} inequality which states that if the
Ricci curvature is bounded from below, then so is the first non-zero
eigenvalue of the Laplace-Beltrami operator. The Laplace-Beltrami operator
differs from the Yang-Mills Hamiltonian in that it lacks the chromo-magnetic
term, but the hope is that in the relevant physical limit the
chromo-electric term dominates the bound. An alternative possibility would
be to generalize the Bochner-Lichn\'{e}rowicz inequality.

Other approaches are based on attempts to solve the Dyson-Schwinger equation
(see for example \cite{SD1} \cite{SD2} \cite{SD3}) on a set of exact
solutions to the classical Yang-Mills theory \cite{Frasca} or on the
ellipticity of the energy operator of cut-off Yang-Mills \cite{Dynin1} \cite%
{Dynin2}.

Once a consistent Euclidean Yang-Mills measure is obtained, the nature of
the mass gap may be found either by computing the distance dependence of the
correlation of two local operators or from the lower bound of the spectrum
in the corresponding Hamiltonian theory. Here I will use the Hamiltonian
approach using the fact that the Hamiltonian may be obtained from the
knowledge of the ground state and the ground state may be obtained from the
measure.

One of the axis directions in the lattice is chosen as the time direction.
Then, recalling that at each step in the projective limit construction one
has a finite-dimensional system, the ground state wave functional $\Psi
_{0}\left( \theta \left( 0\right) \right) $ at a particular configuration $%
\theta \left( 0\right) $ at time zero is obtained by \cite{Rossi} \cite%
{Fradkin}%
\begin{eqnarray}
\left\vert \Psi _{0}\left( \theta \left( 0\right) \right) \right\vert ^{2}
&=&\int d\theta \Psi _{0}^{\ast }\left( \theta \right) \delta \left( \theta
-\theta \left( 0\right) \right) \Psi _{0}\left( \theta \right)  \notag \\
&=&\int d\mu _{\mathcal{A}}\left( \theta \right) \delta \left( \theta
-\theta \left( 0\right) \right)  \label{GS1}
\end{eqnarray}%
where $\mu _{\mathcal{A}}\left( \theta \right) $ is the Euclidean measure
and $\theta $ and $\theta \left( 0\right) $ stand respectively for the set
of group configurations in the edges and for the set of group configurations
in the time-zero slice.

In general the explicit computation of the integral in (\ref{GS1}) is not
easy. However, to study the nature of the mass gap a full calculation of the
ground state wave functional is not required. It uses the interpretation of
elliptic operators as generators of a diffusion process \cite{Friedman} \cite%
{Freidlin1} and, in the limit of small $\beta $, the theory of small
perturbations of dynamical systems \cite{Freidlin2} \cite{Freidlin3}.

The ground state in (\ref{GS1}) may be used to develop the usual Hamiltonian
approach to lattice theory, for which one uses notations similar to those of
Chapter 15 in Ref.\cite{Creutz}, the main difference being that instead of
constructing the Kogut-Susskind Hamiltonian from the Wilson action, one uses
the ground state obtained from the measure. The Hamiltonian will be%
\begin{equation}
H_{g}=\frac{g^{2}\left( \beta \right) }{2\beta }\sum_{l,j,\alpha }\left\{ -%
\frac{\partial }{\partial \theta _{j}^{\alpha }\left( l\right) }%
+L_{j}^{\alpha }\left( l\right) \right\} \left\{ \frac{\partial }{\partial
\theta _{j}^{\alpha }\left( l\right) }+L_{j}^{\alpha }\left( l\right)
\right\}  \label{GS2}
\end{equation}%
The $\theta _{j}^{\alpha }\left( l\right) $'s are the Lie algebra
coordinates of the group element $\exp \left( i\theta _{j}^{\alpha }\left(
l\right) \tau _{\alpha }\right) $ at each edge $l$ of the time-zero slice of
the lattice, the sum is over edges $\left( l\right) $, lattice dimensions $%
\left( j\right) $ and Lie algebra generators $\left( \alpha \right) $. $%
g\left( \beta \right) $ is a coupling constant to be adjusted consistently
to obtain the continuum limit. Recall that from (\ref{II.8}) $\beta
\rightarrow 0$ as the length of the lattice edges ($\frac{a}{2^{k}}$) goes
to zero. 
\begin{equation}
L_{j}^{\alpha }\left( l\right) =-\frac{1}{\Psi _{0}}\frac{\partial \Psi _{0}%
}{\partial \theta _{j}^{\alpha }\left( l\right) }  \label{GS3}
\end{equation}%
which, in particular, implies that the ground state energy $E_{0}$ is
adjusted to zero.

Making the unitary transformation $H_{g}\rightarrow H_{g}^{\prime }=\Psi
_{0}^{-1}H_{g}\Psi _{0}$, the ground state becomes the unit function, all
states are multiplied by $\Psi _{0}^{-1}$ and%
\begin{equation}
-\beta H_{g}^{\prime }=\frac{g^{2}\left( \beta \right) }{2}\sum_{l,j,\alpha }%
\frac{\partial }{\partial \theta _{j}^{\alpha }\left( l\right) }\frac{%
\partial }{\partial \theta _{j}^{\alpha }\left( l\right) }+\sum_{l,j,\alpha
}b_{j}^{\alpha }\left( l\right) \frac{\partial }{\partial \theta
_{j}^{\alpha }\left( l\right) }  \label{GS4}
\end{equation}%
with%
\begin{equation}
b_{j}^{\alpha }\left( l\right) =-g^{2}\left( \beta \right) L_{j}^{\alpha
}\left( l\right) =\frac{g^{2}\left( \beta \right) }{2\Psi _{0}^{2}}\frac{%
\partial \ln \Psi _{0}^{2}}{\partial \theta _{j}^{\alpha }\left( l\right) }
\label{GS5}
\end{equation}%
The second-order elliptic operator in (\ref{GS4}) is the generator of the
diffusion process%
\begin{equation}
d\theta _{j}^{\alpha }\left( l\right) =b_{j}^{\alpha }\left( l\right)
dt+g\left( \beta \right) dW_{j}^{\alpha }\left( l\right)  \label{GS6}
\end{equation}%
with drift $b_{j}^{\alpha }\left( l\right) $ and diffusion coefficient $%
g\left( \beta \right) $. $\Psi _{0}^{2}$ is the invariant measure of this
process. The question of existence of a mass gap for the Hamiltonian $%
H_{g}^{\prime }$ is closely related to principal eigenvalue of the Dirichlet
problem%
\begin{eqnarray}
\beta H_{g}^{\prime }u &=&\lambda u\hspace{1.5cm}\text{in }D  \notag \\
u &=&0\hspace{1.5cm}\text{in }\partial D  \label{GS6a}
\end{eqnarray}%
$D$ being a bounded domain and $\partial D$ its boundary. The principal
eigenvalue $\lambda _{0}$, that is, the smallest positive eigenvalue of $%
\beta H_{g}^{\prime }$ has a stochastic representation \cite{Khasminskii} 
\cite{Freidlin3}%
\begin{equation}
\lambda _{0}=\sup \left\{ \lambda \geq 0;\sup_{\theta \in D}\mathbb{E}%
_{\theta }e^{\lambda \tau }<\infty \right\}  \label{GS6b}
\end{equation}%
$\mathbb{E}_{\theta }$ denoting the expectation value for the process
started from the $\theta $ configuration and $\tau $ the time of first exit
from the domain $D$. The validity of this result hinges on the following
condition

\textbf{(C1)} The drift $b$ and the diffusion matrix coefficient $\sigma $ ($%
g\left( a\right) \delta _{ij}$ in this case) must be uniformly Lipschitz
continuous with exponent $0<\alpha \leq 1$ and $\sigma $ positive definite.

(\ref{GS6b}) is a powerful result which may be used to compute by numerical
means the principal eigenvalue for arbitrary values of $g$ \footnote{%
See for example Ref. \cite{Eleuterio}}. However, a particularly useful
situation is the small noise (small $g$ limit). That the small noise limit
corresponds to the continuum limit of the lattice theory follows from a
consistency argument. Under suitable conditions, to be discussed below, the
small noise limit of the lowest eigenvalue (the mass gap) of the operator $%
\beta H^{\prime }$ is%
\begin{equation}
\beta m\sim \exp \left( -\frac{V}{g^{2}\left( \beta \right) }\right)
\label{GS7}
\end{equation}%
where $V$ is the value of a functional. Hence, for the physical mass gap $m$
to remain fixed when $\beta \rightarrow 0$, it should also be $g\left( \beta
\right) \rightarrow 0$. Therefore the small noise limit is indeed the
continuum limit.

In the small noise limit the mass gap may be obtained from the
Wentzell-Freidlin estimates \cite{Freidlin2} \cite{Freidlin3}. Given a
bounded domain $D$ for the variables $\theta _{j}^{\alpha }\left( l\right) $
define the functional%
\begin{equation}
I_{t_{1},t_{2}}\left( \chi \right) =\frac{1}{2}\int_{t_{1}}^{t_{2}}\left( 
\frac{d\chi }{ds}-b\left( \chi \left( s\right) \right) \right) ^{2}ds
\label{GS8}
\end{equation}%
where $\chi \left( s\in \left[ t_{1},t_{2}\right] \right) $ is a path from
the configuration $\left\{ \theta \right\} $ to the boundary $\partial D$ of
the domain $D$. Then let%
\begin{equation}
I\left( t,\left\{ \theta \right\} ,\partial D\right) =\inf_{\chi
}I_{0,t}\left( \chi \right)  \label{GS9}
\end{equation}%
be the infimum over all continuous paths that starting from the
configuration $\left\{ \theta \right\} $ hit the boundary $\partial D$ in
time less than or equal to $t$. A path is said to be a \textit{neutral path}
if $I\left( t,\left\{ \theta \right\} ,\partial D\right) =0$.

The value of this functional is controlled by the nature of the
deterministic dynamical system%
\begin{equation}
\frac{d\theta _{j}^{\alpha }\left( l\right) }{dt}=b_{j}^{\alpha }\left(
l\right)  \label{GS10}
\end{equation}%
Assume the following aditional condition to be fulfilled:

\textbf{(C2)} There are a number $r$ of $\omega -$limit sets $K_{i}$ of (\ref%
{GS10}) in the domain $D$, with all points in each set $K_{i}$ being
equivalent for the functional $I$, that is, $I\left( t,x,y\right) =0$ if
both $x,y\in K_{i}$ and $b\bullet \nu >0$, $\nu $ being the inward normal to 
$\partial D$.

Then \cite{Friedman} \cite{Freidlin3} with%
\begin{equation}
V_{i}=\inf I\left( t,x,\partial D\right) \hspace{1cm}\text{for }x\in K_{i}
\label{GS11}
\end{equation}%
and%
\begin{eqnarray*}
V_{\ast } &=&\max \left( V_{1},\cdots ,V_{r}\right) \\
V^{\ast } &=&\min \left( V_{1},\cdots ,V_{r}\right)
\end{eqnarray*}%
the lowest non-zero eigenvalue $\lambda _{0}$ satisfies%
\begin{eqnarray*}
\lim_{g\rightarrow 0}\left( -g^{2}\ln \lambda _{0}\left( g\right) \right)
&\leq &V^{\ast } \\
\lim_{g\rightarrow 0}\left( -g^{2}\ln \lambda _{0}\left( g\right) \right)
&\geq &V_{\ast }
\end{eqnarray*}%
In particular if there is only one $V$%
\begin{equation}
\lambda _{0}\left( g\right) =\beta m\left( g\right) \asymp \exp \left( -%
\frac{V}{g^{2}\left( \beta \right) }\right)  \label{GS12}
\end{equation}%
the symbol $\asymp $ meaning logarithmic equivalence in the sense of large
deviation theory. If the drift is the gradient of a function, as in (\ref%
{GS5}), the quasi-potential $V$ is simply obtained from the difference of
the function at the $\omega -$limit set and the minimum at the boundary.

For details on the theory of small perturbations of dynamical systems as
applied to the small $\beta $ limit of lattice theory refer also to \cite%
{Vilela2} where this technique was applied to an approximate ground state
functional. Also \cite{Ludwig1} \cite{Ludwig2} \cite{Ludwig3} \cite{Vilela3}
provide details on how the ground state measure provides a complete
specification of quantum theories both for local and non-local potentials.

Now the existence of a mass gap associated to the Hamiltonian (\ref{GS4}),
obtained from the measure (\ref{II.9}) by (\ref{GS1}), hinges on checking
the above conditions \textbf{(C1)} and \textbf{(C2)}. Inserting (\ref{II.9})
into (\ref{GS1}) one obtains%
\begin{equation}
\left\vert \Psi _{0}\left( g_{l}\left( 0\right) \right) \right\vert
^{2}=\int \prod_{edges}d\mu _{H}(g_{l})\delta \left( g_{l}-g_{l}\left(
0\right) \right) \prod_{plaquettes}\sum_{\lambda \in \Lambda ^{+}}d_{\lambda
}e^{-c\left( \lambda \right) \beta }\chi _{\lambda }\left( g_{p}\right)
\label{GS13}
\end{equation}%
$g_{l}$ being the group element associated to the edges and $g_{p}$ those
associated to the ordered product of group elements around a plaquette, $%
\left\vert \Psi _{0}\left( g_{l}\left( 0\right) \right) \right\vert ^{2}$
being a function only of the group elements on the time slice. For practical
calculations one makes a global lattice gauge fixing in (\ref{GS13}) but for
the present considerations this is not important.

In (\ref{GS13}) the only free variables are the edge variables in the time
slice or, more precicely, the angles of the maximal torus of the group
element associated to the corresponding plaquettes. Smoothness of the heat
kernel implies that the Leibnitz rule for derivation under the integral can
be applied and the drift $b_{j}^{\alpha }\left( l\right) $ in (\ref{GS10})
is also a smooth function. Therefore condition \textbf{(C1)} is satisfied.
As for condition \textbf{(C2)} one knows that the heat kernel satisfies the
following two-sided Gaussian estimate%
\begin{equation}
\frac{1}{\left\vert B\left( e,\beta ^{\frac{1}{2}}\right) \right\vert }%
c_{1}\exp \left( \frac{-d^{2}\left( g\right) }{c_{2}\beta }\right) \leq
K\left( g,\beta \right) \leq \frac{1}{\left\vert B\left( e,\beta ^{\frac{1}{2%
}}\right) \right\vert }c_{3}\exp \left( \frac{-d^{2}\left( g\right) }{%
c_{4}\beta }\right)  \label{GS15}
\end{equation}%
$d\left( g\right) $ being the Carnot-Carath\'{e}odory distance of the group
element $g$ to the identity $e$ and $\left\vert B\left( e,\beta ^{\frac{1}{2}%
}\right) \right\vert $ is the volume of a ball of radius $\beta ^{\frac{1}{2}%
}$ centered at $e$ \cite{Saloff} \cite{Varopoulos}. The estimate (\ref{GS15}%
) holds if and only if

(A) the volume growth has the doubling property%
\begin{equation*}
\forall x\in \mathbb{G},\forall r>0,\left\vert B\left( x,2r\right)
\right\vert \leq c\left\vert B\left( x,r\right) \right\vert
\end{equation*}

(B) there is a constant $\gamma $ such that%
\begin{equation*}
\forall x\in \mathbb{G},\forall r>0,\int_{B\left( x,r\right) }\left\vert
f-Av_{B\left( x,r\right) }f\right\vert ^{2}dx\leq \gamma r^{2}\int_{B\left(
x,2r\right) }\left\vert \nabla f\right\vert ^{2}
\end{equation*}%
$Av_{B\left( x,r\right) }f$ being the average of $f$ over the ball $B\left(
x,r\right) $. In particular if $\mathbb{G}$ is unimodular (B) holds.

For a compact group (A) and (B) being satisfied, the two-sided estimate (\ref%
{GS15}) holds. Therefore the dynamical system (\ref{GS10}) has only one $%
\omega -$limit set, the group identity, and one is in the situation of Eq.(%
\ref{GS12}), $V$ being obtained from the difference of the heat kernel at
the identity and at the boundary of the domain. In conclusion:

\textbf{Theorem: }If $\mathbb{G}$ is a compact group, the Hamiltonian (\ref%
{GS6}) obtained from the heat-kernel measure has a positive mass gap in the $%
\beta \rightarrow 0$ limit, in the sense of Eq.(\ref{GS12}).

The existence of the projective limit\ measure and the projective triplet
made in (I), as well as the characterization of the nature of the mass gap
obtained here, provide a consistent construction of pure Yang-Mills. Of
course, to scale up these results to a full understanding of QCD the role of
fermions as well as of the non-generic strata \cite{Vilela2b} would be
required. In particular to clarify the importance of these strata for the
structure of low-lying excitations.


\begin{thebibliography}{99}
\bibitem{VilelaJMP} R. Vilela Mendes; \textit{An infinite-dimensional
calculus for generalized connections in hypercubic lattices}, J. Math. Phys.
52 (2011) 052304.

\bibitem{Kisynski} J. Kisynski;\textit{\ On the generation of tight measures}%
, Studia Math. 30 (1968) 141-151.

\bibitem{Maurin} K. Maurin; \textit{General eigenfunction expansions and
unitary representations of topological groups}, PWN - Polish Scient. Publ.,
Warszawa 1968.

\bibitem{Ashtekar2} A. Ashtekar and J. Lewandowski; \textit{Differential
geometry on the space of connections via graphs and projective limits}, J.
Geom. Phys. 17 (1995) 191-230.

\bibitem{Ashtekar3} A. Ashtekar and J. Lewandowski; \textit{Projective
techniques and functional integration for gauge theories}, J.Math. Phys. 36
(1995) 2170-2191.

\bibitem{Fleischhack3} C. Fleischhack; \textit{On the support of physical
measures in gauge theories}, arXiv:math-ph/0109030.

\bibitem{Laufer1} M. S. Laufer and P. Orland; \textit{The metric of
Yang-Mills orbit space on the lattice}, Phys.Rev. D88 (2013) 065018

\bibitem{Laufer2} M. S. Laufer; \textit{The Geometry of Lattice-Gauge-Orbit
Space}, Ph. D. Thesis The City University of New York, 2011.

\bibitem{Bochner} S. Bochner; \textit{Vector fields and Ricci curvature},
Bull. Amer. Math. Soc. 52 (1946) 776-797.

\bibitem{Lichne} A. Lichn\'{e}rowicz; \textit{G\'{e}ometrie des groupes de
transformations}, Dunod, Paris 1958.

\bibitem{SD1} R. Alkofer, A. Hauck, L. von Smekal; \textit{Infrared Behavior
of Gluon and Ghost Propagators in Landau Gauge QCD}, Physical Review Letters
79 (1997) 3591-3594.

\bibitem{SD2} V. Gogokhia; \textit{How to demonstrate a possible existence
of a mass gap in QCD}, arXiv:hep-th/0604095v4

\bibitem{SD3} B. Holdom; \textit{Soft asymptotics with mass gap}, Physics
Letters B 728 (2014) 467--471.

\bibitem{Frasca} M. Frasca; \textit{Exact solutions for classical Yang-Mills
fields}, \qquad arXiv:1409.2351

\bibitem{Dynin1} A. Dynin; \textit{Quantum Yang--Mills--Weyl dynamics in the
Schroedinger paradigm,} Russian Journal of Mathematical Physics 21 (2014)
169--188.

\bibitem{Dynin2} A. Dynin; \textit{On the Yang--Mills Mass Gap Problem},
Russian Journal of Mathematical Physics 21 (2014) 326--328.

\bibitem{Rossi} G. C. Rossi and M. Testa; \textit{Ground State Wave Function
from Euclidean Path Integral,} Annals of Physics 148 (1983) 144-167.

\bibitem{Fradkin} E. Fradkin; \textit{Wave functionals for field theories
and path integrals, }Nuclear Physics B389 (1993) 587-600.

\bibitem{Friedman} A. Friedman; \textit{Stochastic differential equations
and applications, vol. 2}, Academic Press, New York 1976.

\bibitem{Freidlin1} M. Freidlin; \textit{Markov processes and differential
equations: Asymptotic problems}, Birkh\"{a}user, Basel 1996.

\bibitem{Freidlin2} A. D. Wentzell and M. I. Freidlin; \textit{On small
random perturbations of dynamical systems}, Russian Math. Surveys 25 (1970)
1--55.

\bibitem{Freidlin3} M. I. Freidlin and A. D. Wentzell; \textit{Random
perturbations of dynamical systems}, Springer, Berlin 2012.

\bibitem{Creutz} M. Creutz; \textit{Quarks, gluons and lattices}, Cambridge
U. P., Cambridge 1983.

\bibitem{Khasminskii} R. Z. Khas'minskii; \textit{On positive solutions of
the equation }$\mathfrak{U}$\textit{u + V \textperiodcentered\ u = 0},
Theory Probab. Appl. 4 (1959) 309--318.

\bibitem{Eleuterio} S. M. Eleut\'{e}rio and R. Vilela Mendes; \textit{%
Numerical predictions from a stochastic model for SU(2) lattice gauge fields}%
, Phys. Lett. B173 (1986) 332-336.

\bibitem{Vilela2} R. Vilela Mendes; \textit{Stochastic processes and the
non-perturbative structure of the QCD vacuum}, Z. Phys. C - Particles and
Fields 54 (1992)\ 273-281.

\bibitem{Ludwig1} S. Albeverio, R. H\o egh-Krohn and L. Streit; \textit{%
Energy Forms, Hamiltonians, and Distorted Brownian Paths}, J. Math. Phys. 18
(1977) 907-917.

\bibitem{Ludwig2} S. Albeverio, R. H\o egh-Krohn and L. Streit; \textit{%
Regularization of Hamiltonians and Processes}, J. Math. Phys. 21 (1980)
1636-1642.

\bibitem{Ludwig3} L. Streit; \textit{Energy forms: Schroedinger theory,
processes}, Physics Reports 77 (1981) 363-375.

\bibitem{Vilela3} R. Vilela Mendes; \textit{Reconstruction of dynamics from
an eigenstate}, J. of Math. Phys. 27 (1986) 178-184.

\bibitem{Saloff} L. Saloff-Coste; \textit{Aspects of Sobolev-type
inequalities}, Cambridge Lect. Notes 289, Cambridge Univ. Press, Cambridge
2002.

\bibitem{Varopoulos} N. Th. Varopoulos.L. Saloff-Coste and T. Coulhon; 
\textit{Analysis and geometry on groups}, Cambridge Tracts on Math. 100,
Cambridge Univ. Press, Cambridge 1992.

\bibitem{Vilela2b} R. Vilela Mendes; \textit{Stratification of the orbit
space in gauge theories. The role of nongeneric strata}, J. Phys. A: Math.
Gen. 37 (2004) 11485-11498.
\end{thebibliography}
\end{document}